\documentclass[structabstract]{aa}
\usepackage{graphicx}
\usepackage{txfonts}
\begin{document}
   \title{The O I] 1641\AA\ line as a probe of symbiotic star winds}

   \author{S. N. Shore\inst{1,2}, G. M. Wahlgren\inst{3,4}}

  \institute{
   Dipartimento di Fisica ``Enrico Fermi'', Universit\`a di Pisa, largo B. Pontecorvo 3, I-56127 Pisa, Italy \email{shore@df.unipi.it}
   \and
   INFN - Sezione di Pisa
   \and
   Catholic University of America, Dept. of Physics, 620 Michigan Ave NE, Washington DC, 20064, USA
   \and
NASA-GSFC, Code 667, Greenbelt, MD, 20771, USA 
             }
              \date{Received ---; accepted ---}
\abstract{}{}{}{}{}
 
  \abstract
   { }
   {The neutral oxygen resonance $\lambda$1302\AA\ line can, if the optical depth is sufficiently high, de-excite by an intercombination transition at $\lambda$1641\AA\ to a metastable state. This has been noted in a number of previous studies but never systematically investigated as a diagnostic of the neutral red giant wind in symbiotic stars and symbiotic-like recurrent novae.}
   {We used archival $IUE$ high resolution, and GHRS and STIS medium and high resolution, spectra to study a sample of symbiotic stars. The integrated fluxes were measured, where possible, for the O I $\lambda$1302\AA\ and O I] $\lambda$1641\AA\ lines. }
   { The intercombination 1641\AA\ line is detected in a substantial number of symbiotic stars with optical depths that give column densities comparable with direct eclipse measures (EG And) and the evolution of the recurrent nova RS Oph 1985 in outburst. In four systems (EG And, Z And, V1016 Cyg, and RR Tel), we find that the O I] variations are strongly correlated with the optical light curve and outburst activity. This transition can also be important for the study of a wide variety of sources in which an ionization-bounded H II region is imbedded in an extensive neutral medium, including active galactic nuclei, and not only for evaluations of extinction. }
   {}


   \maketitle

\section{Introduction}

Symbiotic stars present the unusual situation of a nearly neutral, stable environment centered on a cool giant star, in which a hot source, along with its surrounding ionized region, is imbedded. The radius of the H II region is determined only by the mass gainer's effective temperature and luminosity, which in turn depend only on the accretion rate from the wind (or in the cases where a disk forms, from the flux distribution of the surrounding disk along with that of the underlying star). Since these can be separated using multiwavelength observations, and the incident spectra are simple at ultraviolet (UV) wavelengths (a hot white dwarf and/or an accretion disk continuum and emission line continuum), it is possible to model the formation of the spectrum comparatively easily. This is mainly because the wind of the companion red giant is at nearly its terminal velocity (see Vogel 1991, Pereira et al 1999)  and, even if structured by the orbital motion and hydrodynamic processes related to the accretion (e.g. Dumm et al. 2000; Walder et al. 2008) this happens on a length scale far larger than the gainer and its ionized zone.

In such an environment several radiative processes, not usually encountered under nebular conditions, are observable. Principal among these are fluorescence due to various scattering mechanisms. Accidental resonances account for much of the down-conversion of UV emission to emission in the optical and near infrared. Perhaps the best known are those Fe II and related ions that can be excited by UV resonance transitions of highly-ionized species, e.g. C IV and its coincidence with ground state multiplets of Fe II that de-excite through optical forbidden transitions (Johansson 1983, 1988). Raman scattering (e.g. Schmid 1989), a nearly coincident resonance process that produces broad, down-converted emission lines, is particularly spectacular in the symbiotics, the most notable lines being those of O VI $\lambda\lambda$6825, 7082\AA\ that are produced by the near coincidence of the resonant O VI doublet $\lambda\lambda$1031, 1037\AA\ and H Ly$\beta$. There is, in addition, a process whereby the UV resonance line of a {\it neutral} species can, by virtue of absorption in a surrounding neutral gas, produce both optical and UV emission lines through otherwise inaccessible forbidden transitions. This happens because the ionization potential of several neutral atoms, in particular oxygen, is slightly higher than that of hydrogen and can therefore form in the H II region along with those formed by recombination. Furthermore, resonant absorption by the neutral gas at energies significantly below the ionization limit can, if the optical depth is sufficiently large, lead to emission in alternate channels even in the resonant scattering case. 
 
 The O I spectrum is a case in point. The $\lambda$1302 resonance line is one of the strongest emission features in the spectrum of late-type symbiotics. It forms in the H II region around the degenerate gainer since O I has a slightly higher ionization potential than neutral hydrogen. In addition to the ground state, the O I $^3$S$\rm^o$ $-$ $^3$P ($\lambda\lambda$1302, 1304, 1305\AA) multiplet 5 is connected to two long-lived states through emission at $\lambda$1641\AA\ and $\lambda$2324\AA, both spin forbidden (intercombination) transitions (see Fig. 1), and their associated decay channels to the ground state. One such decay channel, $\lambda$6300 \AA, is well known from terrestrial auroral spectra. Also, the $\lambda$1641 line has been used as a proxy measure of solar activity variability and its effect on the atmosphere (e.g. Bowers et al. 1987, see below). These lines are also well known from planetary nebulae (e.g. Feibelman 1997) and have been discussed in the literature for studies of interstellar extinction in Seyfert galaxies (Grandi 1983) and the determination of the oxygen abundance in cool stars ([O I] $\lambda$6300, Nissen et al. 2002). 
 
 A difficulty presented by any neutral or singly-ionized resonance transitions is that the interstellar medium, possessing the same resonance transitions, is opaque along many lines of sight, especially for distances of several kiloparsecs that are typical of symbiotic and planetary nebular targets. This is exacerbated for cosmological distances where the intervening Ly$\alpha$ forest potentially contaminates the whole redshift range from that of the host galaxy to nearly the local standard of rest. These systems should, therefore, present sufficient line of sight optical depths to produce detectable O I] emission. 
  \begin{figure}
 \centering
 \includegraphics[width=7.5cm, angle=270]{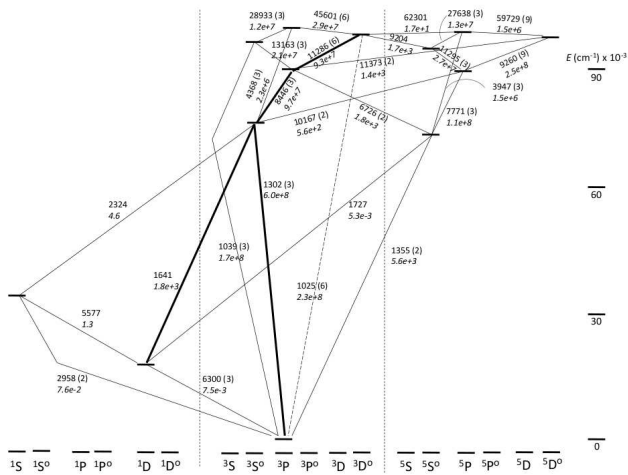}
 \caption{Grotrian diagram for the principal transitions of O I involving all levels up to 10$^5$ cm$^{-1}$.   
 Each multiplet in the figure is labeled with the shortest wavelength (in \AA)
in the multiplet, the number of lines of the multiplet (in parentheses) and the total Einstein $A$ transition probability (in italics, units of s$^{-1}$).   The transition coincident with H Ly$\alpha$ is shown as a long dash, while lines of the prominent decay path and the O I] $\lambda$1641\AA\ line are presented in bold. For clarity, the transition $^5$P (99092 cm$^{-1}$) - $^3$S$^o$ (96225 cm$^{-1}$) ($\lambda$34860 \AA, $A$ = 1.1e01 s$^{-1}$) has not been included. Atomic data are taken from Ralchenko et al. (2008).  }
 \label{spectra}%
 \end{figure}

 In a study of the UV spectra of the recurrent nova RS Oph during its 1985 outburst, Shore \& Aufdenberg (1991) noted the presence of a transient emission line on the red wing of He II $\lambda$1640 relatively early in the outburst and identified this as O I] $\lambda$1641. This line was also identified by Aufdenberg (1993) in the $IUE$ spectrum of RR Tel. In a recent study of the 2006-2009 outburst of the S-type symbiotic star AG Dra, we discussed the variations of the optical spectra, concentrating on the optical Raman features (Shore, Wahlgren, Genovali, et al. 2010). This survey included an examination of archival material as well as optical high-resolution spectra. The absence of the [O I] $\lambda$6300 line was noted but it was suggested that it would be worthwhile checking the existing archive of high resolution UV data for O I] $\lambda$1641. This symbiotic is a special case, having a radial velocity of $-$144 km s$^{-1}$; any resonance line originating from the star is well shifted in wavelength with respect to its ISM components. In this paper we report on our search of the archives for the presence of O I] $\lambda$1641, as well as other O I lines, in the spectra of symbiotic stars. As mentioned above, the O I spectrum originates in the vicinity of the red-giant star. Observations of the O I] line may prove to be useful diagnostics of the red-giant wind and its sources of excitation. Correlation of the UV lines with optical and near-infrared (near-IR) O I lines would therefore enable studies of symbiotic star properties and behavior in the absence of UV spectra.


\section{Observations}

We retrieved all MAST archival spectra for symbiotic stars taken with the {\it International Ultraviolet Explorer} ({\it IUE}) satellite at high resolution (R $\approx$ 10000, large aperture) and {\it Hubble Space Telescope} ($HST$) GHRS and STIS medium resolution (G140M, G160M) spectra. No $HST$ echelle spectra exist for symbiotic stars. For AG Dra, these were supplemented with Telescopio Nazionale Galileo (TNG) high resolution optical spectra. Note was taken of literature sources presenting symbiotic spectra that included O~I lines. Figure 2 presents a sample of O~I] $\lambda$1641 + He~II $\lambda$1640 line profiles for six symbiotics from $HST$/STIS spectra. The high spectral resolution of the STIS instrument clearly shows that the O~I] $\lambda$1641 line can be located either within or outside of the He II $\lambda$1640 profile.

\subsection{Stellar sample}

The available data sets are not of a homogeneous quality, since they were obtained for various scientific purposes, as well as serving as a calibration target. In addition, these data neither represent a thorough nor even sampling of light curves or eruptive events. Therefore, statistics and correlations are directed to the observed occurrence of O~I] $\lambda$1641 and its possible excitation mechanism. Stars for which the line was detected or suspected of being present in the available spectra are the following:
\begin{itemize}
\item Z And: The line is visible in high resolution $IUE$ spectra, its variations are discussed in the following section. The system is both eclipsing (as for CH Cyg) and a very active variable with jet-like outflows having been detected.
\item EG And: Two GHRS spectra show the He II and O I] lines. The forbidden line is strong and easily visible on the lower resolution spectra. There are five STIS pointings, see Crowley et al. (2008) for details. They note the presence of the line but do not study its variations relative to the orbital phase. The main point is that there is almost no variability in the STIS spectra while the He II line is strongly Fe-curtained. The line is present in all spectra taken outside of eclipse, its variation, based on the $IUE$ data, is discussed in the next section.
\item CH Cyg: The UV absorption Fe-curtain spectrum is among the strongest of any symbiotic star.  Although cited by Hack \& Selvelli (1982), the O I] line is weak when present, blended with He II, and visible in two IUE high resolution spectra, SWP8940 (MJD 44365) and SWP10878 (MJD 44596).
\item CI Cyg: There is one GHRS spectrum, showing one of the strongest and most unusual O~I] lines; the intensity relative to He~II is very high. There is one published study (Mikolajewska et al. 2006) that discusses the O I], but it does not discuss the line formation. Possibly present in $IUE$ spectra during the mid-1990s.
\item V1016 Cyg: For the high resolution $IUE$ spectra, many are saturated at He II and partially mask any weaker emission lines, but the O I] line is apparent in spectrum SWP05612 and possibility detected in other spectra. Only a single G140L STIS spectrum is available from the archives, the resolution of which is insufficient to detect the O I]. The line variations are discussed in the next section.
\item V1329 Cyg: The O~I] line is possibly present at a level just above the approximate continuum noise level for several $IUE$ spectra, especially in SWP29816.
\item AG Dra: O I] is strong, one STIS spectrum shows this perfectly. Weakly present in the red wing of He II $\lambda$1640 in the $IUE$ spectrum SWP25444. The O I] line is, however, too weak to measure in the archival $IUE$ spectra and there is only one {\it STIS} spectrum (2003 Apr. 19, see Shore et al. 2010), for which the O I] $\lambda$1641 flux is $8.13\times 10^{-14}$ erg s$^{-1}$cm$^{-2}$ and the (O I $\lambda$1302/O I] $\lambda$1641) ratio is 10.2. Mikolajewska et al. (1995) propose a moderately small inclination and no eclipses. 
 \item RW Hya: Two GHRS and one STIS low resolution spectra show only a hint of O~I], mainly as a redward-extended wing on He II that could mask the unresolved line.
\item SY Mus: High resolution $IUE$ spectra (including SWP14236) show the O I] line.
\item AG Peg: Two GHRS spectra give a possible indication of the line, but it is clearly weak. Possibly present in the $IUE$ spectrum SWP37420, the shortest exposure available.
\item RX Pup: There is one STIS spectrum but no detectable emission at O I]. An $IUE$ spectrum (SWP14240) shows the line.
\item HM Sge: Only high resolution $IUE$ are available, and there is a possibility that these show the O I] line.
\item RR Tel: The most studied of the symbiotic stars in this sample, spectra were taken with GHRS and STIS as a standard for calibration of wavelengths and comparisons between instruments. The O I] line is very strong. The line is also strong in $IUE$ spectra.
\item KX TrA: Weakly present in the $IUE$ spectra SWP38741, SWP38742.
\end{itemize}

\subsection{Notes and correlations}

For the stars AS210, S190, R Aqr, AE Ara, T CrB, BF Cyg, and AX Per the O I] line was either absent or too weak to measure in the available data, in all cases only $IUE$ spectra exist. Based on the available spectra mentioned, two-thirds of the symbiotic stars have shown the presence of O~I] $\lambda$1641. Among the remaining classical symbiotics from Kenyon (1986), Allen (1984), and Belcy\'nski et al (2000), Y CrA, V443 Her, BX Mon, AR Pav, CL Sco, HK Sco, CL Sco, and AS296 have only low resolution {\it IUE} archival spectra, and are therefore not useful for detecting O~I] $\lambda$1641.

    \begin{figure}
   \centering
   \includegraphics[width=9cm]{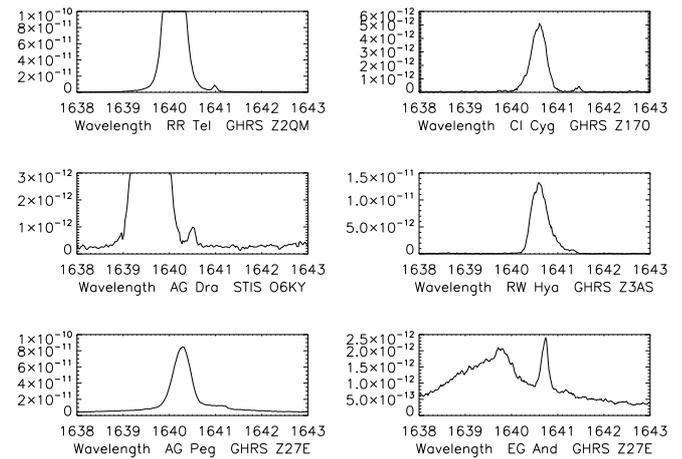}
   \caption{Sample of O I] $\lambda$1641 + He II $\lambda$1640\AA\ profiles in a subset of symbiotic stars observed with the GHRS and STIS, medium resolution spectra.  The HST program is indicated, monochromatic flux unit is erg s$^{-1}$cm$^{-1}$\AA$^{-1}$}
              \label{spectra}%
    \end{figure}

Although not measured in the previous studies (especially the outburst analysis by Shore et al. (1996)), the O I] line is detected in {\it IUE} high dispersion spectra of the 1985 outburst of the symbiotic-like recurrent nova RS Oph. In Fig. 3 we show three early spectra, from 10 days (SWP 25248), 25 days (SWP 25290), and 30 days (SWP 25328) after optical maximum. The O I] line is clearly detected in only one of these, SWP 25290, which was obtained during the broad-line phase of the permitted lines and when the inferred neutral column density was about $10^{23}$cm$^{-2}$ based on the Fe-curtain absorption (derived from the narrow line components of C IV, see Shore et al. (1996)). Its radial velocity is consistent with that of the binary system, $v_{\rm rad} = -10$ km s$^{-1}$. To our knowledge, the line O I $\lambda$1.12895 $\mu$m has only been identified in symbiotic and related star spectra by Evans et al. (2007), where it was recorded in post outburst spectra of RS Oph.

    \begin{figure}
   \centering
   \includegraphics[width=8.5cm]{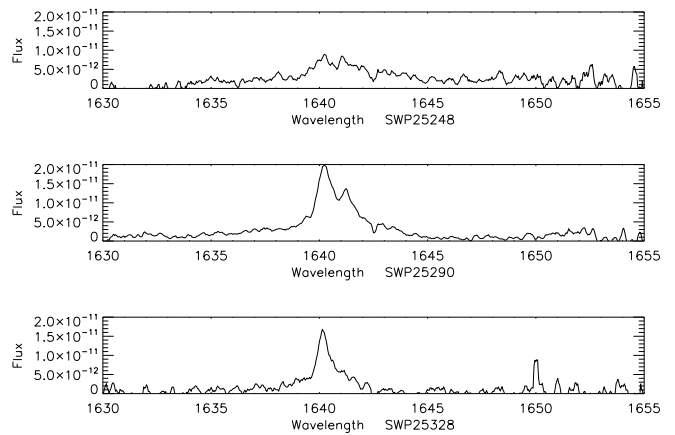}
   \caption{The He II $\lambda$1640 and O I] $\lambda$1641\AA\ emission in {\it IUE} high dispersion spectra of RS Oph during the early stage of the 1985 outburst (see text for details)}
              \label{spectra}%
    \end{figure}

We show four examples of long term variability of this line in Fig. 4 for EG And, Fig. 5 for V1016 Cyg, Fig.6 and Fig.7 for RR Tel, and Fig. 8 for Z And. The precipitous drop
in the RR Tel flux corresponds to a step in the visual magnitude light curve, obtained from the AAVSO website. Similarly, the spike in the Z And O I]
flux corresponds to a spike in the visual magnitude light curve. The AAVSO visual magnitude light curve for V1016 Cyg shows a weak correlation with peaks
in the O I] flux in Fig. 5. The sparser EG And data in Fig. 4 is more difficult to correlate with its light curve; however, the amplitude of the O I] flux
is similar to the visual magnitude amplitude over a period that is similar to the 482 day period of its (orbit/pulsation). This latter is an eclipsing system in which the line of sight optical depth is so large, due to the absorption line and Rayleigh scattering opacities, as to obscure the entire region, including the O I] line. For non-eclipsing systems, the optical depth appears never to be so large that the O I], which is always optically thin, is not seen. As with the Raman feature, the line is formed close to the ionized-neutral interface in the red giant wind and for low inclination systems should always be visible.
   \begin{figure}
   \centering
   \includegraphics[width=8cm]{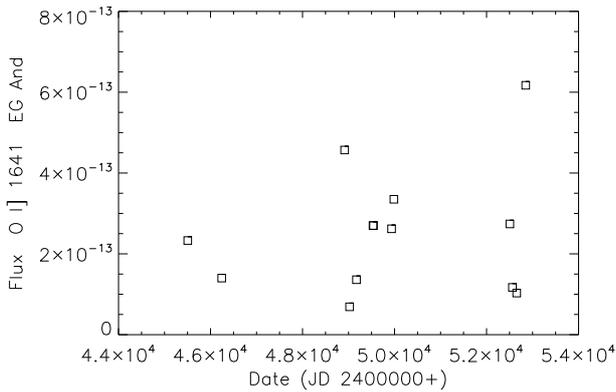}
   \caption{EG And: variations of the integrated flux of the O I] $\lambda$1641\AA\ ({\it IUE, GHRS}, and {\it STIS} spectra) }
              \label{spectra}%
    \end{figure}
    
Parimucha et al. (2002) have shown that the fluxes for resonance lines of ionized species (C III-C IV, N III-N V, etc.) in V1016 Cyg showed a well defined minimum at around MJD 47000. It is possible that the O I] line, which shows an almost lightcurve, may reach minimum strength slightly earlier. 

\begin{figure}
   \centering
   \includegraphics[width=8cm]{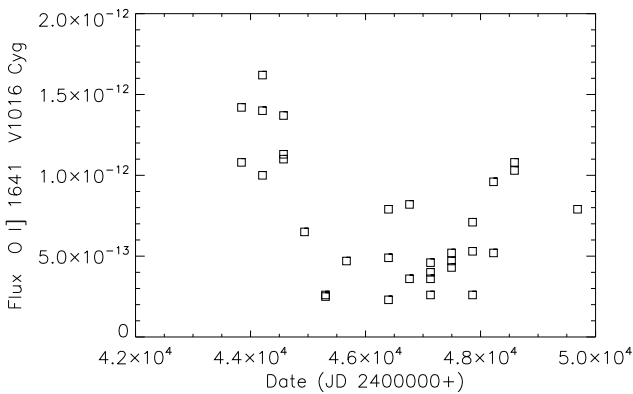}
   \caption{V1016 Cyg: variations of the O I] $\lambda$1641\AA\ line flux, high resolution large aperture {\it IUE} data. }
              \label{spectra}%
    \end{figure}

The secular development of the O I] strength in RR Tel is the same as the slow decline of the $V$ magnitude, according to the AAVSO light curves, for the entire period of the $IUE$ measurements. In addition, since in this system the stellar radial velocity suffices to displace the O I $\lambda$1302\AA\ line from within the interstellar absorption, it is possible to study the long term variation of the $\lambda$1302\AA\ to $\lambda$1641\AA\ line flux ratio, shown in Fig. 7. There is an apparently asymptotic trend with F($\lambda$1302)/F($\lambda$1641)$\approx$ 0.4 at late times in the IUE data set. This flux ratio agrees with the STIS observations from almost a decade later.

    \begin{figure}
   \centering
   \includegraphics[width=8.5cm]{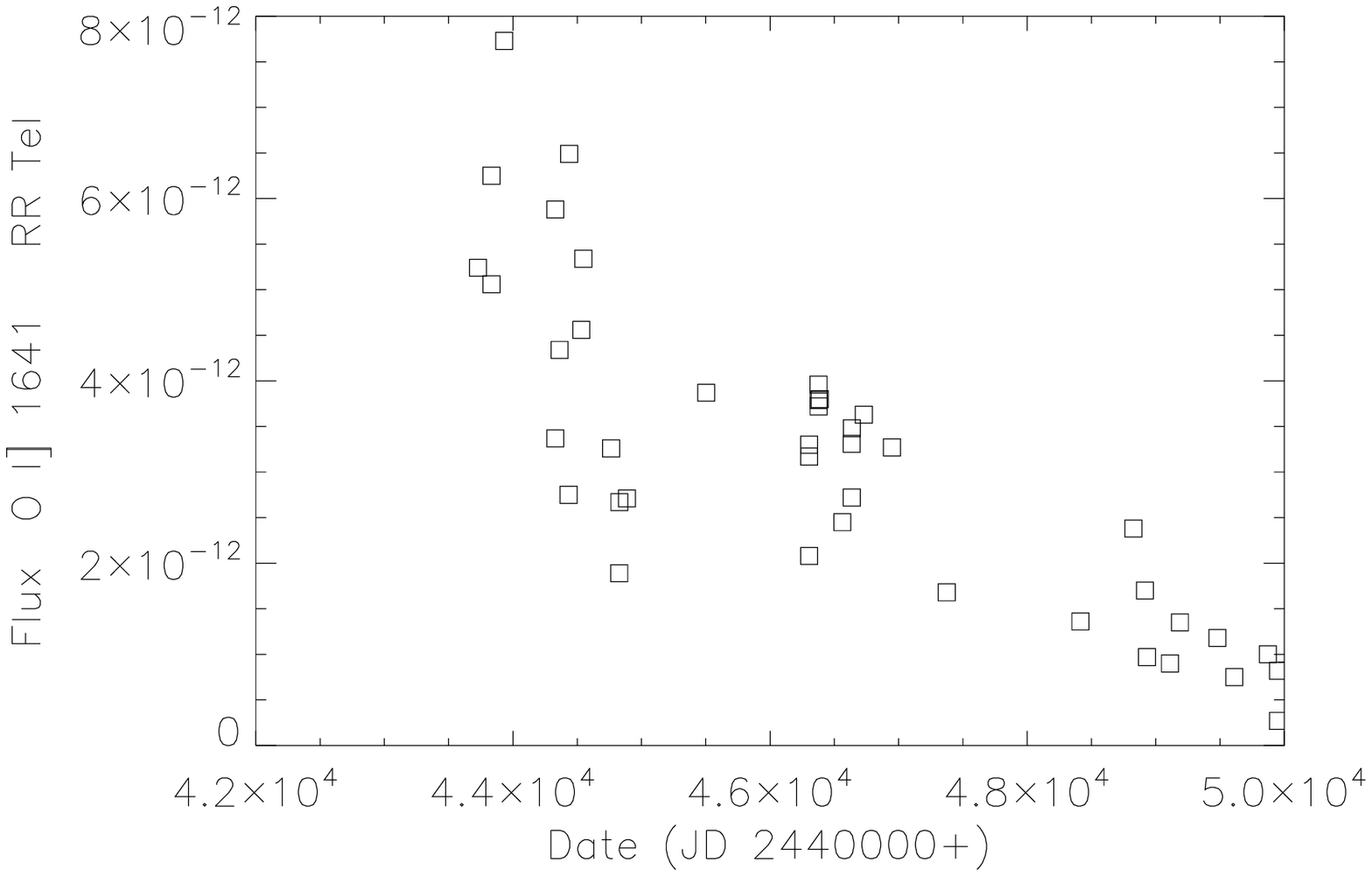}
   \caption{RR Tel: variations of the O I] 1641\AA\ line flux, high resolution large aperture $IUE$ data. }
              \label{spectra}%
    \end{figure}
    
    \begin{figure}
   \centering
   \includegraphics[width=8.5cm]{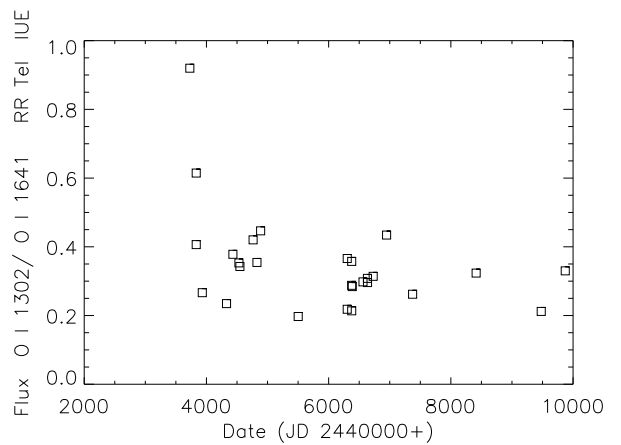}
   \caption{RR Tel: variations of the O I $\lambda$1302/O I] $\lambda$1641 line ratio (uncorrected for extinction), high resolution large aperture $IUE$ data. }
              \label{spectra}%
    \end{figure}

    \begin{figure}
   \centering
   \includegraphics[width=8.5cm]{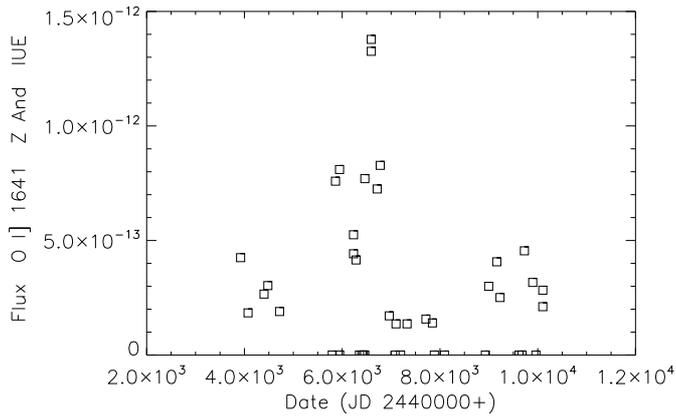}
   \caption{Z And: variations of the O I] $\lambda$1641\AA\ line flux, high resolution large aperture $IUE$ data. }
              \label{spectra}%
    \end{figure}
      The integrated flux variations of the $\lambda$1641\AA\ line for Z And (Fig. 8) shows a strong correlation with the long term optical variations (based on the AAVSO archive), especially the strong outburst between MJD 46000 and 47000. In the figure, the zero flux values are the conjunctions when it is heavily obscured (according to Friedjung et al. 2010) that were otherwise well exposed (not, as in several symbiotic stars, over- or underexposures that we have ignored in the analysis). 
      
The {\it HST} data set includes extended wavelength coverage at high spectral resolution and signal-to-noise for the stars
AG Dra, EG And, and RR Tel. This allows for a search of multiple lines from O I for the purpose of investigating
emission line excitation mechanisms. STIS spectra for AG Dra extend from the vacuum UV to the red. Emission
is observed for $\lambda\lambda$1302, 1304, 1305, 1355, 1358, 1641.
The $\lambda$1304 line is nearly entirely removed by saturated absorption from the ISM feature Si II $\lambda$1304.370, while
the O I $\lambda$1302 emission line is impinged upon by its saturated ISM counterpart (as an example we show the 1300\AA\ region of AG Dra in Fig. 9).  Several lines ($\lambda\lambda$2324, 8446, 9204, 9260) may have marginal detections, and no evidence
exists for the presence of others ($\lambda\lambda$5577, 6300, 6363, 6391). The $\lambda$2324 line is suspect due
to broadening of the 2324 \AA\ feature (C II + O I).

    \begin{figure}
   \centering
   \includegraphics[width=8.5cm]{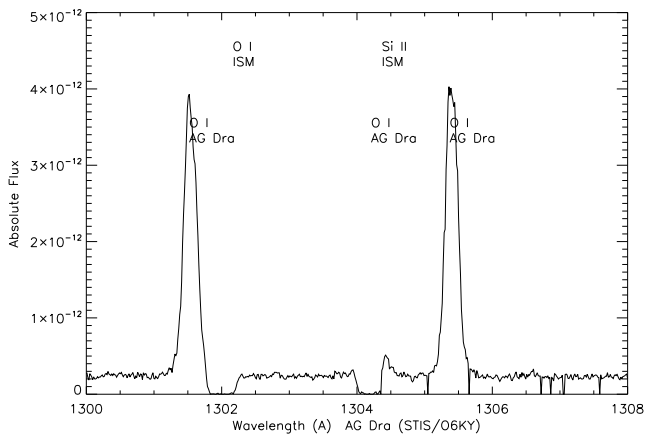}
   \caption{AG Dra: {\it STIS} medium resolution spectrum (program O6KY) showing the interstellar and stellar lines at O I $\lambda\lambda$1302, 1304, and 1306 and Si II $\lambda$1304.  The stellar radial velocity, about -150 km s$^{-1}$, is sufficient to almost completely displace the resonance line from within the ISM profile but a chance coincidence obscures the $\lambda$1304 line.  The lefthand side of the annotation indicates the rest wavelength.}%
    \end{figure}

For EG And the $HST$ data are limited to UV wavelengths, showing emission for $\lambda\lambda$1302,1304,1305,1355, and 1358. Emission
in O I $\lambda$2324.738, if present, is blended with the emission feature C II $\lambda$2324.69, which is one of five
lines that comprise the C II UV2 2s$^2$ 2p $^2$P$^o$ - 2s 2p$^2$ $^4$P multiplet, all of which are found in emission. O I $\lambda$1727.106 does not appear to be present. Emission is not found for lines at the longer wavelengths ($\lambda\lambda$2958, 2972)
as the continuum flux of the cool star, the increased number of absorption lines, and the lower transition probabilities render these
lines difficult to detect. The RR Tel data set clearly shows emission from O I $\lambda\lambda$1302, 1304, 1305, 1641, 5577, 6300, 6363.
The detection of $\lambda$2324 line is complicated by blending with C II, with the lines of this multiplet not being in proportion to
their relative $gA$ values. Strong emission at $\lambda$1025 is not evident (see below).

The O I] $\lambda$1641 emission appears to be positively correlated with the O VI Raman emission
at $\lambda\lambda$6825, 7082. Based on ORFEUS spectra, Schmid et al. (1999) compared emission from the O VI $\lambda\lambda$1032, 1038 \AA\
resonance doublet with the Raman scattering features for six symbiotic stars. Of these six,
five have UV spectra that can be searched for O I]. We find that the strongest O VI Raman emitters (AG Dra, RR Tel) are also the strongest
O I] emitters, while three others (AG Peg, Z And, V1016 Cyg) present both weak O I] and weak O VI Raman lines. The O VI FUV doublet has two contributors, the white dwarf wind and surrounding ionized cavity within the red giant wind, that are Raman scattered by Ly$\beta$ absorption by the H I in the wind. The O I] line samples the same warm neutral H wind zone. Crowley et al. (2008), for instance, find that this lies about 10$^{13}$cm (a small fraction of the semimajor axis of the system) from the white dwarf and it is in the cooler zone that both oxygen and hydrogen are neutrals. Since the O I 1302\AA\ line arises from recombination in the H II region, whose boundary is virtually the same for the two species, the 1641\AA\ line probes the same region as the Raman feature. During outburst events, for instance for Z And around MJD 46700 (Fig. 7), the O I] line notably and briefly strengthened and when the He II line was hidden by a sufficient optical depth in the Fe-curtain absorption the O I] was also obscured.
     
\section{Discussion}
     
Several fluorescence processes can lead to the population of levels that will ultimately lead to emission of O I] 1641 \AA. Coincidence of H Ly$\beta$
$\lambda$1025.722 with O I $\lambda$1025.762 will populate the O I 3d $^3$D$^o$ (97488 cm$^{-1}$) level from the ground level.
The dominant decay chain (according to their Einstein transition probabilities, see Fig. 1) from the 3d $^3$D$^o$ level ($\lambda\lambda$11285 and 8446)
will populate the 3s $^3$S$^o$ (76794 cm$^{-1}$) level, which subsequently decays through three channels ($\lambda\lambda$1302, 1641, 2324),
two of which are commonly detected in symbiotic star spectra. In addition to the Ly$\beta$ O I pumping, McMurry \& Jordan (2000) identified CO emission fluorescently-excited by O I UV 2 resonance line emission near $\lambda$1302 in the UV spectrum of $\alpha$ Tau.

A second pumping mechanism for the O I 76794 cm$^{-1}$ level is He II $\lambda$1640 for those stars which have a broad He II line. This is
evident for RR Tel and EG And, and less so for RW Hya and AG Peg, in Fig. 2. Other possible pumping mechanisms that might lead to
population enhancement of this O I level are:
1) H Ly$\epsilon$ $\lambda$937.803 \AA\ coincident with O I $\lambda$937.841 to pump the O I 106765 cm$^{-1}$ level, which can decay
to the 76794 cm $^{-1}$ level through the chain $\lambda\lambda$14110, 4368 \AA, among others.
2) H Ly$_6$ $\lambda$930.748 and He II $\lambda$930.342 can pump the O I 8s $^3$S$^o$ 107497 cm$^{-1}$ level, which decays
to the 76794 cm$^{-1}$ level through the chains $\lambda\lambda$12790, 4368 or $\lambda\lambda$5298, 8446.
3) C II $\lambda$2324.69 emission is coincident with O I $\lambda$2324.738 and can pump the O I 76794 cm$^{-1}$ level from the metastable O I level 33792 cm$^{-1}$.
The five lines of C II multiplet 2 are seen in emission in a number of symbiotics and symbiotic novae (RR Tel).
Direct excitation by the resonance line should, however, be more effective in symbiotics, as in the terrestrial case since the Ly lines are so optically thick and the illumination is from the companion, not {\it in situ} from the chromosphere (there will, of course, be a contribution from the spectrum of the late-type component but this is small compared to that from white dwarf environment). 

Population of the O I 76794 cm$^{-1}$ level via electron recombination is possible through additional decay chains.
Spectral observations at infrared wavelengths may offer a means of determining the dominant excitation mechanisms by detecting
other emission lines. The number of lines from the O I spectrum that have been observed in astronomical targets, in particular symbiotic stars
and novae, are few. Common UV lines detected include transitions at wavelengths $\lambda\lambda$1302, 1304, 1305, 1355, 1358, 1641.
At optical wavelengths $\lambda\lambda$6300, 6363 are found in planetary nebula spectra  with $\lambda$6300 commonly used for abundance analysis
in cool stars. For near-IR wavelengths, detections, or suspicions of detections, have been mentioned for 8446 \AA\ in AG Dra (Iijima et al. 1987),
and $\lambda$11289 (Evans et al. 2007). Absorption lines at $\lambda\lambda$7771, 7773, 7774 are commonly used in abundance analysis in a variety of stars.
There is also the curious appearance of an undiscussed weak emission line near 2.9 $\mu$m (Schild, Boyle \& Schmid 1992) in spectra
of several symbiotics. Conspicuous by their absence from discussion and published spectra are lines of large transition probability
($\lambda\lambda$4368, 9204, 9260 for example) and small transition probability ($\lambda\lambda$1727, 2324, 2958, 2972). A full accounting of O I lines for any target
would be useful for determining the excitation conditions and better enable the physical modeling.

The importance of the O I] $\lambda$1641 line for the symbiotics is as a possible tool for as long as ultraviolet spectroscopy is available. Oddly,
although this line has been included in a number of identification lists at high resolution, it has never been exploited as
a diagnostic for symbiotics or related systems. It has, however, been noted as a contributor to the energetics of AGN when the
O I resonance line is sufficiently optically thick. Grandi (1983), in discussing reddening determinations for AGN using the resonance and Bowen fluorescence O I UV2 lines ($\lambda\lambda$1302, 1304 vs $\lambda$8446) noted that the line ratio O I] $\lambda$1641 (UV146) to the resonance multiplet is often unusually large, given the branching ratio. This can be accounted for by a large enough optical depth to strongly self-absorb the ground state lines. With a Ly$\alpha$ optical depth as large as $10^6$ the reduction in $\lambda$1302 is sufficient to produce an integrated flux of only a factor of 2 larger than the forbidden transition. The inhomogeneous regions around the central engine often show such large opacities while still permitting observation of the nucleus along a given sight line. More recently, the chromospheric O I spectrum has been rediscussed for a few main-sequence and evolved F, G, and K stars by Koncewicz (2005) and Koncewicz \& Jordan (2007). There is another mode to produce the O I] $\lambda$1641 emission, the coincidence of Ly$\beta$ $\lambda$1025.72 and the O I resonance line (UV4) at $\lambda$1025.77 that pumps the $^3$D$^o$ 97488 cm$^{-1}$ level , which then decays through the $\lambda\lambda$11286, 8446, 1641, 6300 chain.
 
To date, however, most of the literature deals with the O I
lines in the context of planetary -- specifically, terrestrial -- atmospheric structure and
composition. Atomic oxygen forms in excited states by dissociative 
collisions between O$_2$ and electrons. These transitions
have been used for studying the oxygen abundance and temperature
structure of the troposphere in a number of papers, e.g. Meier \&
Conway (1985), and Conway et al. (1988). Doering \& Gulcicek
(1989) include the $\lambda\lambda$1355, 1358 lines. Since the O I] transition
is always optically thin and absent in the reflected solar spectrum
this transition probes almost the entire terrestrial stratosphere
and ionosphere. The branching ratio (Garstang 1961, Erdman \&
Zipf 1986) is O I] $\lambda$1641/O I $\lambda$1302 = 5.1$\times$10$^{-6}$ with an uncertainty
of 30\% . The most recent compilation, Wiese et al. (1996), gives
$A$($\lambda$1641)/$A$($\lambda$1302) = 5.4$\times$10$^{-6}$. Following Grandi (1983), based
on escape probability formalism (Kwan \& Krolik 1981), we can
estimate the required column density in the resonance line. The
observed branching ratio is $\approx$ 1 for all systems in which the O
I] 1641\AA\ line is detected in our survey, the implied optical depth
for $\lambda$1302 is $\tau_{\lambda1302} \approx 7\times 10^{-14}f v_{50}^{-1}
N_O \approx 2\times10^5$, where $f$ is the oscillator strength and $v_{50}$ is the wind velocity in km s$^{-1}$, that  for a solar
O/H ratio ($5\times 10^{-4}$, see Asplund et al. 2009) gives a column density N$_H$ $\approx$ 10$^{23}$ cm$^{-2}$ for the
neutral absorption region.  This is the same order of magnitude as
the column density in absorption required to explain the narrow
UV emission line variations during the early RS Oph outburst
(Shore et al. 1996) and similar to that derived by Crowley et
al. (2008) from eclipse spectra of EG And. Using the length scale
from the photoionization modeling in Crowley et al. (2008), who
obtain a standoff distance for the neutral region from the white
dwarf in EG And of about 10$^{13}$ cm, gives a characteristic
number density of about 10$^{10}$ cm$^{-3}$. For a wind velocity of 50 km
s$^{-1}$ and using R$_{13}$ =(R/10$^{13}$ cm) gives an estimate of the mass loss rate for the red
giant of $\approx$ 10$^{-6}$ R$^2$$_{13}$ M$_{\sun}$ yr$^{-1}$. This estimate is different between systems (there
are several with lower branching ratios, others with higher, and in many cases the stellar radial velocity is not sufficient to displace
the $\lambda$1302 line from within the interstellar absorption. The optical depth expected for the red-giant wind in the FUV O I
doublet, combined with the opacity of the Ly$\beta$ transition, suggest
that this is not a dominant mechanism in producing the $\lambda$1641
line and that the O I opacity suffices.

\begin{acknowledgements}

We thank J. P. Aufdenberg, K. Genovali, J. Mikolajewska, C. Rossi, and R. Viotti, and the (anonymous) referee for valuable discussions and suggestions.   The $IUE$ and the $HST$ GHRS and STIS spectra were obtained from the MAST archive of STScI and archival visual photometric data were provided 
by the AAVSO. GMW acknowledges support from NASA Grant NNG06GJ29G.
 
\end{acknowledgements}

\end{document}